\documentclass[prc,aps]{revtex4}
\usepackage{epsfig,array}
\begin{document}
\newcommand{\be}{\begin{equation}}
\newcommand{\ee}{\end{equation}}
\newcommand{\bea}{\begin{eqnarray}}
\newcommand{\eea}{\end{eqnarray}}
\newcommand{\f}{\frac}  
\newcommand{\la}{\lambda}
\newcommand{\ve}{\varepsilon}
\newcommand{\ep}{\epsilon}
\newcommand{\da}{\downarrow}
\newcommand{\up}{\uparrow}
\newcommand{\V}{{\cal V}}
\newcommand{\ovl}{\overline}
\newcommand{\Ga}{\Gamma}
\newcommand{\ga}{\gamma}
\newcommand{\bra}{\langle}
\newcommand{\ket}{\rangle}
\newcommand{\eff}{_{\rm{eff}}}
\newcommand{\av}{{\rm{av}}}
\newcommand{\fl}{{\rm{fl}}}
\newcommand{\ina}{{\rm{in}}}
\newcommand{\wt}{\widetilde}
\newcommand{\ov}{\overline}
\newcommand{\G}{{\cal G}}
\newcommand{\Ha}{{\cal H}}
\newcommand{\sig}{\sigma}
\title{Energy averages and fluctuations in the decay out of
superdeformed bands}
\author{A.~J.~Sargeant$^1$, M.~S.~Hussein$^1$, M.~P.~Pato$^1$ and M.~Ueda$^2$}
\affiliation{$^1$Nuclear Theory and Elementary Particle Phenomenology 
Group, Instituto de F\'\i sica, Universidade de S\~{a}o Paulo,
Caixa Postal 66318, 05315-970 S\~{a}o Paulo, SP, Brazil}
\affiliation{$^2$Institute of Physics, University of Tsukuba, Ten-noudai 1-1-1,
Tsukuba, Ibaraki, 305-8571, Japan}
\date{October, 4 2002}
\begin{abstract}
We derive analytic formulae for the energy average (including the 
energy average of the fluctuation contribution) and variance of the 
intraband decay intensity 
of a superdeformed band. Our results may be expressed in terms of three
dimensionless variables: $\Ga^\da/\Ga_S$, $\Ga_N/d$ and
$\Ga_N/(\Ga_S+\Ga^\da)$. Here $\Ga^\da$ is the spreading width for the 
mixing of a superdeformed (SD) state $|0\ket$ with the normally deformed 
(ND) states $|Q\ket$ whose spin is the same as $|0\ket$'s.
The $|Q\ket$ have mean lever spacing $d$ and mean electromagnetic decay width 
$\Ga_N$ whilst $|0\ket$ has electromagnetic decay width $\Ga_S$.

The average decay intensity may be expressed solely in terms of the 
variables $\Ga^\da/\Ga_S$ and $\Ga_N/d$ or, analogously to 
statistical nuclear reaction theory, in terms
of the transmission coefficients $T_0(E)$ and $T_N$ describing
transmission from the $|Q\ket$ to the SD band via $|0\ket$ and
to lower ND states. 

The variance of the decay intensity, in analogy with
Ericson's theory of cross section fluctuations, depends 
on an additional variable, the correlation length 
$\Ga_N/(\Ga_S+\Ga^\da)=\f{d}{2\pi}T_N/(\Ga_S+\Ga^\da)$.
This suggests that analysis of an experimentally determined variance could 
yield the mean level spacing $d$ as does analysis of the 
cross section autocorrelation function in compound nucleus reactions.

We compare our results with those of Gu and Weidenm\"uller.
\end{abstract}
\maketitle
\section{Introduction}
The basic feature to be explained in the decay-out of superdeformed (SD)  
rotational bands 
\cite{La 02,Lo 00,Dewald:2001hs,Krucken:2001we} is that the 
intensity of the collective $\ga$--rays emitted during the cascade down
an SD band remains constant until a certain spin is reached 
where-after it drops to zero within a few transitions. 
The sharp drop in intensity is believed to arise from mixing of the SD states 
with normally deformed (ND) states of identical spin \cite{Vi 90}. 
The model of Refs.~\cite{Sh 92,Sh 93,Sh 00,Yo 00} 
attributes the suddenness of the decay-out to the spin dependence of the 
barrier separating the SD and ND minima of the deformation potential.
Refs. \cite{Ab 99,Krucken:vr,Sargeant:2001aq} discuss the effect 
of the chaoticity of the ND states on the decay-out.

In the present paper we derive analytic formulae for the energy average 
(including the energy average of the fluctuation contribution) and variance 
of the intraband decay intensity of a superdeformed band
in terms of variables which usefully describe the decay-out.
\cite{Dewald:2001hs,Krucken:2001we,Gu:1999bv,We 98}. 
We achieve this using the MIT 
approach to statistical nuclear reaction theory \cite{Fe 92,Ka 73,Er 63}. 
The MIT approach tackles calculation
of the fluctuation cross section and other moments of the $S$-matrix  
by directly calculating averages of fluctuating functions of energy. 

In agreement with Gu and Weidenm\"uller \cite{Gu:1999bv} (GW)
we find that average of the total intraband decay
intensity can be written as a function of the dimensionless variables
$\Ga^\da/\Ga_S$ and $\Ga_N/d$ where
$\Ga^\da$ is the spreading width for the 
mixing of an SD state with normally ND states of the
same spin, $d$ is the mean level spacing of the latter and 
$\Ga_S$ $(\Ga_N)$ are the electromagnetic decay widths of the SD (ND) 
states. Our formula for the variance of the total intraband decay intensity,
in addition to the two dimensionless variables just mentioned,
depends on the dimensionless variable $\Ga_N/(\Ga_S+\Ga^\da)$. This
additional variable is analogous to the correlation width of the 
Ericson's theory of cross section fluctuations. Its appearance suggests that
measurement of the variance of the decay intensity could yield the mean level 
spacing $d$.

The paper is organised in the following manner. In Section~\ref{KKM}
we express the average decay intensity as an average term plus
the average of a fluctuation term. In Section~\ref{intensity} the expression
for the average decay intensity obtained in Section~\ref{KKM}
is evaluated approximately. In Section~\ref{variance} we calculate the
variance of the intensity within the same approximation scheme as 
Section~\ref{intensity}. In Section~\ref{comp} we interpret our
results by analogy with statistical nuclear reaction theory, expressing our
results in terms of transmission coefficients and compare our results
with those of GW. In particular we suggest a reason why GW did not observe 
the variable $\Ga_N/(\Ga_S+\Ga^\da)$ in their calculation of the variance. 
Finally in Section~\ref{conclusions} we make some concluding remarks 
including mention of the limitations of our results.
\section{Energy averages and fluctuations}
\label{KKM}
The total intraband decay intensity has the form 
\cite{Wa 84,Gu:1999bv,Sargeant:2001aq}
\be
I_\ina=\left(2\pi\Ga_S\right)^{-1}\int_{-\infty}^{\infty}
dE|A_{00}(E)|^2,
\label{I}
\ee
where the intraband decay amplitude is given by
\be
A_{00}(E)=\ga_S\langle \wt{0}|G(E)|0\rangle\ga_S.
\label{A}
\ee
Here $\ga_S$ is the electromagnetic decay amplitude of 
superdeformed state $|0\ket$ defined such that 
$\Ga_S=\ga_S^2$. We assume that the amplitude feeding $|0\ket$
is also given by $\ga_S$. The total Green's function, $G$, is given by
\be
G(E)=(E-H)^{-1},
\label{G}
\ee
where the full nuclear Hamiltonian is denoted by $H$ and has a non-Hermitian
part which accounts for coupling to the electromagnetic field 
[see Eqs. (\ref{PQ}), (\ref{HPP}) and (\ref{HQQ}) below].
The tilde is used to indicate the dual state or adjoint 
\cite{Fe 92} of $|0\ket$.  

In what follows we employ the optical background representation introduced
Kawai, Kerman and McVoy \cite{Ka 73}. These authors investigated 
fluctuation cross sections using a representation of the $S$-matrix in which 
the background $S$-matrix is chosen to be the energy average of the 
$S$-matrix itself, that is, the $S$-matrix corresponding to the optical 
potential. Here we use the same idea to decompose thedecay amplitude, 
Eq.~(\ref{A}), into the sum of its Lorentzian energy average (Lorentzian 
energy averaging interval $I$), 
\be\nonumber
\ov{A_{00}(E)}\equiv A_{00}(E+iI/2), 
\ee
plus a fluctuating part. 

We proceed by introducing Feshbach's projection operators \cite{Fe 92}
\be\label{PQ}
P=|\wt{0}\ket\bra 0|\hspace{.2cm} \mbox{and}\hspace{.2cm}Q=1-P.
\ee
Let us introduce the notation $\G=PGP$ for the effective Green's 
function in the $P$-space and let $H_{PP}=PHP$, $H_{PQ}=PHQ$ etc.
Using the techniques of Ref. \cite{Ka 73} we obtain
\be\label{Geff}
\G=\ov{\G}+\G^\fl,
\ee
where the average effective Green's function $\ov{\G}=\G(E+iI/2)$ is given by
\be\label{Gav}
\ov{\G}=\left[E-\ov{\Ha}\right]^{-1}
\ee
and the average effective Hamiltonian is given by
\be
\ov{\Ha}=H_{PP}+\ov{W}_{PP},
\label{Hav}
\ee
where coupling to the eliminated $Q$-space is accounted for by
\be
\ov{W}_{PP}=H_{PQ}\f{1}{E-H_{QQ}+iI/2}H_{QP}=\f{-2i}{I}V_{PQ}V_{QP}
\label{WPP}
\ee
and the energy dependent coupling potential, $V$, introduced in 
Ref.~\cite{Ka 73} is defined by
\bea\nonumber
V_{PQ}&=&H_{PQ}\sqrt{\f{iI/2}{E-H_{QQ}+iI/2}},
\\\nonumber
V_{QP}&=&\sqrt{\f{iI/2}{E-H_{QQ}+iI/2}}H_{QP}.
\eea
The fluctuating part of the effective Green's function, $\G$, is 
given by
\be\label{Gfl}
\G^\fl=\ov{\G}V_{PQ}\f{1}{E-H_{QQ}-W_{QQ}}V_{QP}\ov{\G},
\ee
where coupling back to the $P$-space is accounted for by
\be
W_{QQ}=V_{QP}\ov{\G}V_{PQ}.
\label{WQQ}
\ee
By construction $\ov{\G^\fl}=0$ in so much as $\ov{\G}$ is unchanged by
reaveraging, that is if $\ov{\ov{\G}}=\ov{\G}$.

The corresponding decomposition of Eq.~(\ref{A}) for the transition amplitude 
is
\be
A_{00}(E)=\ov{A_{00}(E)}+A_{00}^\fl(E),
\label{2A}
\ee
where the energy average of the transition amplitude is
\be
\ov{A_{00}}=\ga_S\ov{\G}_{00}\ga_S
\label{Aav}
\ee 
and the fluctuating part of the transition amplitude is
\be
A_{00}^\fl=\ga_S\G^\fl_{00}\ga_S.
\label{Afl}
\ee
Thus the average of Eq.~(\ref{I}) for the relative intensity may be written 
as the incoherent sum
\be
\ov{I_\ina}=I_\ina^\av+\ov{I_\ina^\fl},
\label{ovI}
\ee
where
\be
I_\ina^\av=\ov{I_\ina^\av}=\left(2\pi\Ga_S\right)^{-1}
\int_{-\infty}^{\infty}dE
\left|\ov{A_{00}(E)}\right|^2
\label{Iav}
\ee
and
\be
\ov{I_\ina^\fl}=\left(2\pi\Ga_S\right)^{-1}
\int_{-\infty}^{\infty}dE
\ov{\left|A_{00}^\fl(E)\right|^2}.
\label{ovIfl}
\ee

Up to this point no assumptions have been made except that the 
transition amplitude can be written in the form of Eq.~(\ref{A}). As will
be made clear in Section~\ref{intensity} the meaning of this assumption is
that $|0\ket$ is a doorway for the decay from the SD band to the ND states
and vice versa. In the manipulations subsequent to Eq.~(\ref{A}) we have
put the average intensity in a form consisting of a background term 
coming from the smooth energy dependence of the doorway plus a term resulting
from fluctuations on this background. 

The representation we have used is particular suitable for approximation when 
the ND states are overlapping. In Section~\ref{intensity} we evaluate 
Eqs.~(\ref{ovI}-\ref{ovIfl}) for the average decay intensity, $\ov{I_\ina}$, 
assuming that this is the case. In Section~\ref{variance} we calculate the 
variance which describes the way in which $I_\ina$ fluctuates about 
$\ov{I_\ina}$.
\section{Average Decay Intensity}
\label{intensity}
Let us assume that $H_{PP}$ satisfies the eigenvalue equation
\be\label{HPP}
H_{PP}|0\ket=(E_0-i\Ga_S/2)|0\ket
\ee 
and $H_{QQ}$ 
\be\label{HQQ}
H_{QQ}|Q\ket=(E_Q-i\Ga_N/2)|Q\ket.
\ee 
Here, $E_0$ denotes the energy of SD state $|0\ket$, $\Ga_S$ it's 
electromagnetic width for decay to the next lowest state in the SD band, 
$E_Q$ ($Q=1,...,N$) the energy of the $N$ ND states $|Q\ket$ with the same 
spin as $|0\ket$ and $\Ga_N$ the common electromagnetic width of the $|Q\ket$ 
for decay to ND states of lower spin. 
 
Further, let us write the matrix element of $W_{PP}$, Eq.~(\ref{WPP}), as
\bea\label{W00}
\ov{W}_{00}&=&\bra 0|\ov{W}_{PP}|0\ket=\Delta^\da-i\Ga^\da/2.
\eea
Here, $\Delta^\da=\mbox{Re} \ov{W}_{00}$ is an energy shift which we ignore
and $\Ga^\da=-2\mbox{Im} \ov{W}_{00}$. 
Combining these definitions with Eqs.~(\ref{Gav}), (\ref{Hav}) and (\ref{Aav})
the average of the transition amplitude can be written
\be\label{2Aav}
\ov{A_{00}}=\f{\Ga_S}{E-E_0+i(\Ga_S+\Ga^\da)/2}.
\ee
We see that Eq.~(\ref{2Aav}) exhibits the structure of an isolated
doorway resonance. The doorway $|0\ket$ has an escape width $\Ga_S$ for 
decay to the SD state with next lower spin and a spreading width $\Ga^\da$ for
decay to the ND states with the same spin which are reached
by tunnelling through the barrier separating the SD and ND wells.
The doorway structure of Eq.~(\ref{2Aav}) is due to the assumption that the 
transition amplitude can be written as in Eq.~(\ref{A}). The most general 
expression for the transition amplitude has the form 
$A_{ab}=\ga_{ab}+\sum_{cc'}\ga_{ac}\bra c|G|c'\ket\ga_{c'b}$ where
$\ga_{ac}$ describes the coupling of channels $a$ and $c$.
In our doorway model the $c$ and $c'$
stand for $|0\ket$ or $|Q\ket$, $Q=1,...,N$,
ie. $\sum_{c}|c\ket\bra c|=P+Q$, and $a$ and $b$ denote channels
the (electromagnetic) coupling to which is taken is accounted for by
the non-Hermitian part of $H$ [Eqs. (\ref{HPP}) and (\ref{HQQ})],
that is they denote the SD state above $|0\ket$, the SD below $|0\ket$ and 
ND states whose spin is different from that of $|0\ket$. The direct coupling of
channels $a$ and $b$, $\ga_{ab}$, is taken to be zero.

In order to evaluate $A^\fl_{00}$ it is useful to introduce eigenvectors and 
eigenvalues of the operator $H_{QQ}+W_{QQ}$ defined by
\be\label{q}
\left(H_{QQ}+W_{QQ}\right)|q\ket=(E_q-i\Ga_q/2)|q\ket,
\hspace{.5cm}q=1,...,N.
\ee
Then from Eq.~(\ref{Afl}) and Eq.~(\ref{Gfl}) for $\G^\fl$ we obtain
\bea\nonumber
A^\fl_{00}&=&\ga_S\ov{\G}_{00}
\sum_q\f{\bra \wt{0}|V_{PQ}|q\ket\bra \wt{q}|V_{QP}|0\ket}
{E-E_q+i\Ga_q/2}\ov{\G}_{00}\ga_S
\\&=&\ov{A_{00}}\sum_q
\f{g_{0q}g_{q0}/\Ga_S}{E-E_q+i\Ga_q/2}\ov{A_{00}},
\label{2Afl}
\eea
where
\bea\label{g}
g_{0q}&=&\bra \wt{0}|V_{PQ}|q\ket,
\\g_{q0}&=&\bra \wt{q}|V_{QP}|0\ket.
\eea

We now employ some statistical assumptions which are frequently 
used in statistical nuclear reaction theory \cite{Fe 92} to derive
an analytic formula for the decay intensity $\ov{I_\ina}$.
We shall assume that the Lorentzian and box energy averages and the average
over the label, $q$, are all approximately
equal, ie., that for a suitable function $f_q(E)$ of $q$ and $E$ 
\be
\ov{f_q(E)}=f_q(E+iI/2)
\approx\f{1}{\Delta E}\int_{E_0+\Delta E/2}^{E_0+\Delta E/2}dE'f_q(E')
\approx\f{1}{N}\sum_{q=1}^Nf_q(E).
\label{avequiv}
\ee
The width of the box average, $\Delta E$, is related to the 
width the Lorentzian energy average by $\Delta E\approx\pi I/2$ and to 
the mean spacing, $d$, of the $N$ ND states by $\Delta E\approx Nd$.
This approximation is good as long as $\Ga/d\gg1$ [see Eq.~(\ref{Ga}) below
for the definition of $\Ga$]. Within these assumptions 
we see from Eq.~(\ref{WPP}) that the $g_{0q}$ and 
$g_{q0}$ are related to $\ov{W}_{00}$ by
\bea
\ov{W}_{00}&=&\f{-2i}{I}\sum_qg_{0q}g_{q0}
\approx\f{-\pi i}{\Delta E}\sum_qg_{0q}g_{q0}\approx
\f{-\pi i}{d}\ov{g_{0q}g_{q0}}.
\label{2W00}
\eea
Thus the spreading with is given by
\be\label{Gada}
\Ga^\da=\f{2\pi}{d}\mbox{Re}\hspace{1mm}\ov{g_{0q}g_{q0}}
\ee
and the energy shift
by
\be\label{shift}
\Delta^\da=\f{\pi}{d}\mbox{Im}\hspace{1mm}\ov{g_{0q}g_{q0}}.
\ee

From Eq.~(\ref{2Afl}) we can calculate the amplitude autocorrelation function 
\bea
\ov{A_{00}^\fl(E){A_{00}^\fl(E')}^*}&=&
\f{1}{{\Ga_S}^2}\hspace{1mm}{\ov{A_{00}(E)}}^2\hspace{1mm}
\ov{\sum_{qq'}\f{g_{0q}g_{q0}g_{0q'}^*g_{q'0}^*}
{\left(E-E_q+i\Ga_q/2\right)\left(E'-E_{q'}-i\Ga_{q'}/2\right)}}
\ov{{A_{00}(E')}^*}^2
\label{Aflauto}
\\&=&
\f{1}{{\Ga_S}^2}\hspace{1mm}{\ov{A_{00}(\ov{\ve}+\ve/2)}}^2\hspace{1mm}
\ov{\sum_{qq'}\f{g_{0q}g_{q0}g_{0q'}^*g_{q'0}^*}
{\left(\ov{\ve}+\ve/2-E_q+i\Ga_q/2\right)
\left(\ov{\ve}-\ve/2-E_{q'}-i\Ga_{q'}/2\right)}}
\ov{{A_{00}(\ov{\ve}-\ve/2)}^*}^2,
\label{1Aflauto}
\eea
where we have made the variable changes $\ve=E-E'$ and $\ov{\ve}=(E+E')/2$.
Consider the middle factor in Eq.~(\ref{1Aflauto}) which
we anticipate is a function of $\ve$ only
\bea\label{a}
a(\ve)=\ov{\sum_{qq'}\f{g_{0q}g_{q0}g_{0q'}^*g_{q'0}^*}
{\left(\ov{\ve}+\ve/2-E_q+i\Ga_q/2\right)
\left(\ov{\ve}-\ve/2-E_{q'}-i\Ga_{q'}/2\right)}}.
\eea
We interpret the energy average in Eq.~(\ref{a}) to be an average over
$\ov{\ve}$. Employing a box energy average
\bea\label{aa}
a(\ve)\approx\f{1}{\Delta E}{\sum_{qq'}
\int_{E_0-\Delta E/2}^{E_0+\Delta E/2}d\ov{\ve}
\f{g_{0q}g_{q0}g_{0q'}^*g_{q'0}^*}
{\left(\ov{\ve}+\ve/2-E_q+i\Ga_q/2\right)
\left(\ov{\ve}-\ve/2-E_{q'}-i\Ga_{q'}/2\right)}}.
\eea
If $\Delta E$ is large compared to the $\Ga_q$ but small enough for
the $E_q$ and $\Ga_q$ to be treated as constants then we may extend the
limits of integration to $\pm\infty$ and perform the integral using the
calculus of residues to obtain
\bea\label{aaa}
a(\ve)\approx\f{2\pi i}{\Delta E}
{\sum_{qq'}\f{g_{0q}g_{q0}g_{0q'}^*g_{q'0}^*}
{\ve+E_{q'}-E_q+i(\Ga_{q'}+\Ga_q)/2}}.
\eea
Assuming that the phases of the $g_{0q}$ and the $g_{q0}$ are randomly 
distributed as a function of $q$ the double sum in Eq.~(\ref{aaa}) collapses to
a single sum giving
\bea\label{aaaa}
a(\ve)&\approx&\f{2\pi i}{\Delta E}
\sum_{q}\f{|g_{0q}g_{q0}|^2}
{\ve+i\Ga_q}.
\eea
Then employing the definition of the average given by Eq.~(\ref{avequiv})
\bea
a(\ve)&\approx&\f{2\pi i}{d}
\ov{\left[\f{|g_{0q}g_{q0}|^2}{\ve+i\Ga_q}\right]}.
\label{1a}
\eea
Assuming that the average of a ratio is equal to the ratio of 
the averages we get
\bea\label{2a}a(\ve)&\approx&\f{{\Ga^\da}^2}{2\pi}
\f{id}{\ve+i\Ga},
\eea
with $\Ga^\da$ and $\Ga$ introduced according to Eq.~(\ref{ovgg}) and
Eq.~(\ref{Ga}) below. The introduction of $\Ga^\da$ is based on the 
assumption that
\bea\label{ovgg}
\ov{\left|g_{0q}g_{q0}\right|^2}
\approx 2\left|\ov{g_{0q}g_{q0}}\right|^2
\approx 2\left[\f{\Ga^\da d}{2\pi}\right]^2.
\eea
The factor of 2 which appears in the first manipulation in Eq.~(\ref{ovgg})
accounts for the self-correlation (present since the entrance and exit channel
are both $|0\ket$) and is equal to the elastic enhancement 
factor for compound elastic scattering in the overlapping resonance region.
In the second manipulation in Eq.~(\ref{ovgg})
we have again ignored the energy shift 
[see Eqs.~(\ref{Gada}) and (\ref{shift})].

Eq. (\ref{2a}) also introduces the correlation width
\bea\label{Ga}
\Ga&\approx&\ov{\Ga_q}
\\\label{1Ga}&=&-2\mbox{Im}\ov{\bra \wt{q}|H_{QQ}+W_{QQ}|q\ket}
\\\label{2Ga}&=&\Ga_N+\Ga^\up
\\\label{3Ga}&\approx&\Ga_N,
\eea
where we used Eq.~(\ref{HQQ}) for the electromagnetic width of the $|Q\ket$ 
and introduced
\bea\nonumber\Ga^\up
&=&-2\mbox{Im}\ov{\bra \wt{q}|W_{QQ}|q\ket}
\\\nonumber&=&-2\mbox{Im}\f{\ov{g_{q0}g_{0q}}}{E-E_0+i(\Ga_S+\Ga^\da)/2}
\\&=&\f{\Ga^\da d}{2\pi}
\f{(\Ga_S+\Ga^\da)}{(E-E_0)^2+(\Ga_S+\Ga^\da)^2/4},
\label{Gaup}
\eea
which is the width for their decay back to $|0\ket$.
The approximations represented by Eqs.~(\ref{Ga}) and (\ref{3Ga})
will be discussed in Section~\ref{comp}. Using Eq. (\ref{2a}) and 
approximation (\ref{3Ga}) we finally obtain
\bea
\ov{A_{00}^\fl(E){A_{00}^\fl(E')}^*}
&\approx&2\left(2\pi\Ga_N/d\right)^{-1}
(\Ga^\da/\Ga_S)^2\hspace{2mm}
{\ov{A_{00}(E)}}^2\hspace{2mm}\f{i\Ga_N}{E-E'+i\Ga_N}\ov{{A_{00}(E')}^*}^2.
\label{2Aflauto}
\eea
When $E'=E$ this reduces to  
\be
\ov{\left|A_{00}^\fl\right|^2}=2\left(2\pi\Ga_N/d\right)^{-1}
\f{{\Ga_S}^2{\Ga^\da}^2}
{\left[(E-E_0)^2+(\Ga_S+\Ga^\da)^2/4\right]^2},
\label{Aflsqu}
\ee
which is the average of the fluctuation contribution to the transition 
intensity.

The integrals in Eqs.~(\ref{Iav}) and (\ref{ovIfl}) for the average and 
fluctuation contributions to the total decay intensity may now be carried out
using the calculus of residues. Substituting Eq.~(\ref{2Aav}) into 
Eq.~(\ref{Iav}) we obtain 
\bea
I_\ina^\av=\f{1}{1+\Ga^\da/\Ga_S}.
\label{2Iav}
\eea
Eq.~(\ref{2Iav}) is identical with the equivalent result in GW (see
also Ref.~\cite{We 98}).
Substituting Eq.~(\ref{Aflsqu}) into Eq.~(\ref{ovIfl}) we obtain
\bea
\ov{I_\ina^\fl}&=&2\left(\pi\Ga_N/d\right)^{-1}
\f{\left(\Ga^\da/\Ga_S\right)^2}
{\left(1+\Ga^\da/\Ga_S\right)^3}
\label{2ovIfl}
\\&=&2\left(\pi\Ga_N/d\right)^{-1}
I_\ina^\av\left(1-I_\ina^\av\right)^2.
\label{3ovIfl}
\eea
for the average fluctuation contribution to the average decay intensity.
\begin{figure}[ht]
\centerline{\epsfig{figure=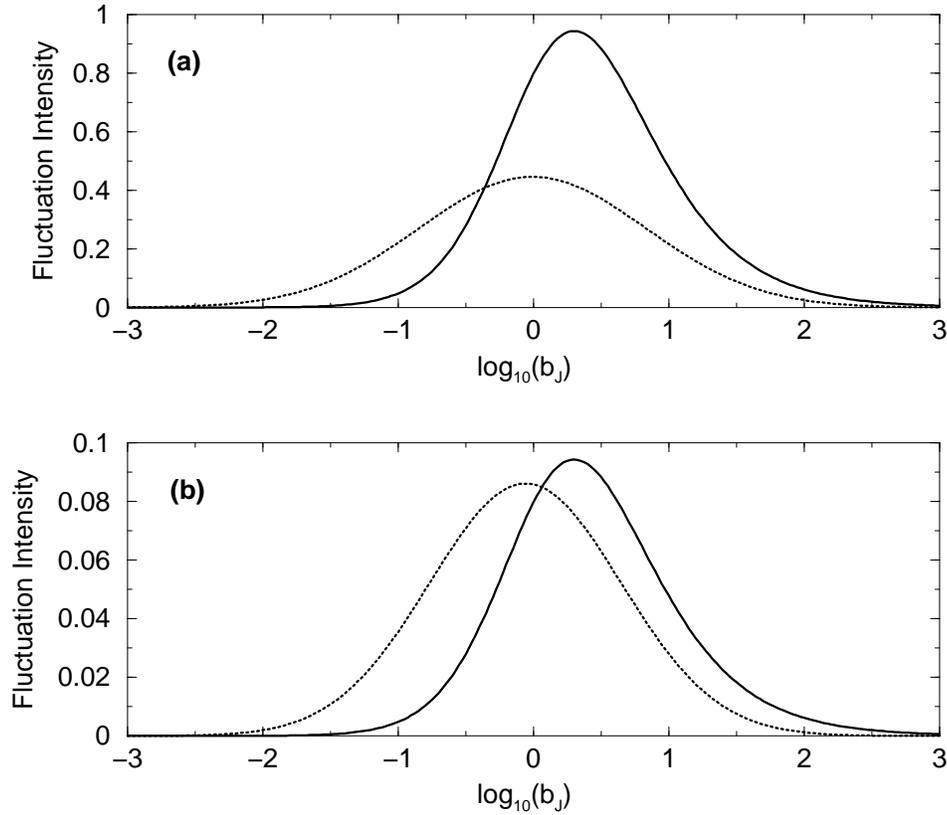,width=.7\textwidth}}
\caption{Average of the fluctuation contribution to the intraband intensity 
$\ov{I_\ina^\fl}$~vs.~$\log_{10}(b_J)$ where $b_J\equiv\Ga^\da/\Ga_S$. The solid 
lines were calculated using Eq.~(\ref{2ovIfl}) and the dotted lines by GW's 
fit formula, Eq.~(\ref{gufit}). The variable $\Ga_N/d$ took the value 0.1 for
graph (a) and 1 for graph (b).}
\label{flucfig1}
\end{figure}

Eq.~(\ref{2ovIfl}) for $\ov{I_\ina^\fl}$ is plotted in Fig.~\ref{flucfig1} 
and for comparison we have also plotted a fit formula which was 
obtained by GW,
\bea
\ov{I_\ina^\fl}&=&\left[1-0.9139\left(\Ga_N/d\right)^{0.2172}
\right]
\exp\left\{-\f{\left[0.4343\ln\left(\f{\Ga^\da}{\Ga_S}\right)
-0.45\left(\f{\Ga_N}{d}\right)^{-0.1303}\right]^2}
{\left(\Ga_N/d\right)^{-0.1477}}\right\}.
\label{gufit}
\eea
Qualitative agreement is seen between the two formulae. 
Note that Eq.~(\ref{gufit}) yields a negative intensity for 
$\Ga_N/d>1.51$ which excludes its use in this regime.
Our result which is only strictly valid when $\Ga_N/d\gg 1$ 
is simply inversely proportional to $\Ga_N/d$.
The exact result of GW for $\ov{I_\ina^\fl}$ [Eq.~(24) in GW]
can be used for any $\Ga_N/d$ also decreases monotonically with
increasing $\Ga_N/d$.

The dependence of $\ov{I_\ina^\fl}$ (and that of $I_\ina^\av$) 
on $\Ga^\da/\Ga_S$ results from the resonant doorway energy 
dependence of the decay amplitude $\ov{A_{00}(E)}$ [Eq.~(\ref{2Aav})]. 
This energy dependence also manifests itself in the average of the fluctuation
contribution to the transition intensity $\ov{|A^\fl_{00}(E)|^2}$ 
[Eq.~(\ref{Aflsqu})]. GW include precisely the same energy dependence
in their calculation by use of an energy dependent transmission coefficient to 
describe decay to the SD band [see Eq.~(\ref{T0}) below and the discussion in 
Subsection~\ref{gu}]. This is the reason for our qualitative agreement with
GW concerning $\ov{I_\ina}$. 
\section{Variance of the decay intensity}
\label{variance}
The error incurred in making the energy average is given by
\be\nonumber
\Delta I_\ina=I_\ina-\ov{I_\ina}
=I_\ina^\fl-\ov{I_\ina^\fl}+2\mbox{Re}
\int_{-\infty}^{\infty}dE \ov{A_{00}(E)}{A^\fl_{00}(E)}^*.
\ee
The average of the error vanishes: $\ov{\Delta I_\ina}=0$.
A measure of the dispersion of the calculated $I_\ina$ is given by the
variance 
\be\label{var}
\ov{\left(\Delta I_\ina\right)^2}
=\ov{\left(I_\ina-\ov{I_\ina}\right)^2}.
\ee
In order to evaluate $\ov{\left(\Delta I_\ina\right)^2}$
the averages indicated in Eq.~(\ref{var}) must be performed 
{\em before} the integration which appears in the definition of $I_\ina$,
Eq.~(\ref{I}). We obtain
\be\ov{\left(\Delta I_\ina\right)^2}
\approx\left(2\pi\Ga_S\right)^{-2}
\int_{-\infty}^{\infty}dE\int_{-\infty}^{\infty}dE'\left\{{\hspace{2mm}
\left|\ov{A_{00}^\fl(E){A_{00}^\fl(E')}^*}\right|^2
+2\mbox{Re}\hspace{1mm}
\ov{{A_{00}(E)}^*}
\hspace{1mm}\ov{A_{00}^\fl(E){A_{00}^\fl(E')}^*}
\hspace{1mm}\ov{A_{00}(E')}}\right\}.
\label{2var}
\ee
In deriving Eq.~(\ref{2var}) we have used 
$\ov{A_{00}^\fl(E)}$=$\ov{A_{00}^\fl(E)A_{00}^\fl(E')}$=$0$,
$\ov{A_{00}^\fl(E){A_{00}^\fl(E')}^*}$$\ne$$0$ and
$\ov{A_{00}^\fl(E){A_{00}^\fl(E)}^*A_{00}^\fl(E'){A_{00}^\fl(E')}^*}$=
$\ov{\left|A_{00}^\fl(E)\right|^2}
\hspace{2mm}\ov{\left|A_{00}^\fl(E')\right|^2}$
+$\left|\ov{A_{00}^\fl(E){A_{00}^\fl(E')}^*}\right|^2$ 
and we have assumed that averages of terms containing odd powers 
of $A^\fl_{00}$ vanish.

Substituting Eqs.~(\ref{2Aflauto}) and (\ref{2Aav})
in Eq.~(\ref{2var}), making the changes of integration variable
\bea\label{x}
E-E_0&=&(\Ga_S+\Ga^\da)x/2
\\\label{x'}E'-E_0&=&(\Ga_S+\Ga^\da)x'/2 
\eea
and using Eq.~(\ref{2Iav}) and (\ref{2ovIfl}) for
$\ov{I_\ina^\av}$ and $\ov{I_\ina^\fl}$ we are able to write
$\ov{\left(\Delta I_\ina\right)^2}$ in the form
\be\label{3var}
\ov{\left(\Delta I_\ina\right)^2}=
{\ov{I_\ina^\fl}}^2f_1\left(\xi\right)
+2I_\ina^\av\ov{I_\ina^\fl}f_2\left(\xi\right),
\ee
where the variable $\xi$ is defined by
\bea\label{xi}
\xi&\equiv&\f{\Ga_S+\Ga^\da}{\Ga_N}
\\\label{2xi}&=&\f{\Ga_S}{\Ga_N}(1+\Ga^\da/\Ga_S)=\f{\Ga_S}{\Ga_N}
{I_\ina^\av}^{-1}
\\\label{3xi}&=&\f{\Ga^\da}{\Ga_N}(1+\Ga_S/\Ga^\da)^{-1}
=\f{\Ga^\da}{\Ga_N}(1-I_\ina^\av).
\eea
The functions $f_1$ and $f_2$ which have been introduced in Eq~(\ref{3var}) 
are given by
\bea\label{1f1}
f_1(\xi)&=&\left(\f{4}{\pi\xi}\right)^2
\int_{-\infty}^{\infty}\f{dx}{\left[x^2+1\right]^2}
\int_{-\infty}^{\infty}\f{dx'}{[\left(x-x'\right)^2+4/\xi^2]
\left[x'^2+1\right]^2}
\\\nonumber&=&\f{-8}{\pi^2\xi}\mbox{Im}
\int_{-\infty}^{\infty}\f{dx}{\left[x^2+1\right]^2}
\int_{-\infty}^{\infty}\f{dx'}{\left[x-x'+2i/\xi\right]
\left[x'+i\right]^2\left[x'-i\right]^2}
\\&=&\f{-4}{\pi\xi}\mbox{Im}
\int_{-\infty}^{\infty}\f{dx}{\left[x^2+1\right]^2}
\left(\f{1}{x+i(2/\xi+1)}+\f{i}{\left[x+i(2/\xi+1)\right]^2}\right)
\label{f1}
\eea
and
\bea\label{1f2}
f_2(\xi)&=&\f{4}{\pi^2\xi}\mbox{Re}\hspace{2mm}i
\int_{-\infty}^{\infty}\f{dx}{\left[x-i\right]\left[x+i\right]^2}
\int_{-\infty}^{\infty}\f{dx'}
{\left[x-x'+2i/\xi\right]\left[x'-i\right]^2\left[x'+i\right]}
\\&=&\f{-2}{\pi\xi}\mbox{Re}
\int_{-\infty}^{\infty}\f{dx}{\left[x-i\right]\left[x+i\right]^2
\left[x+i(2/\xi+1)\right]}.
\label{f2}
\eea
\begin{figure}[ht]
\centerline{\epsfig{figure=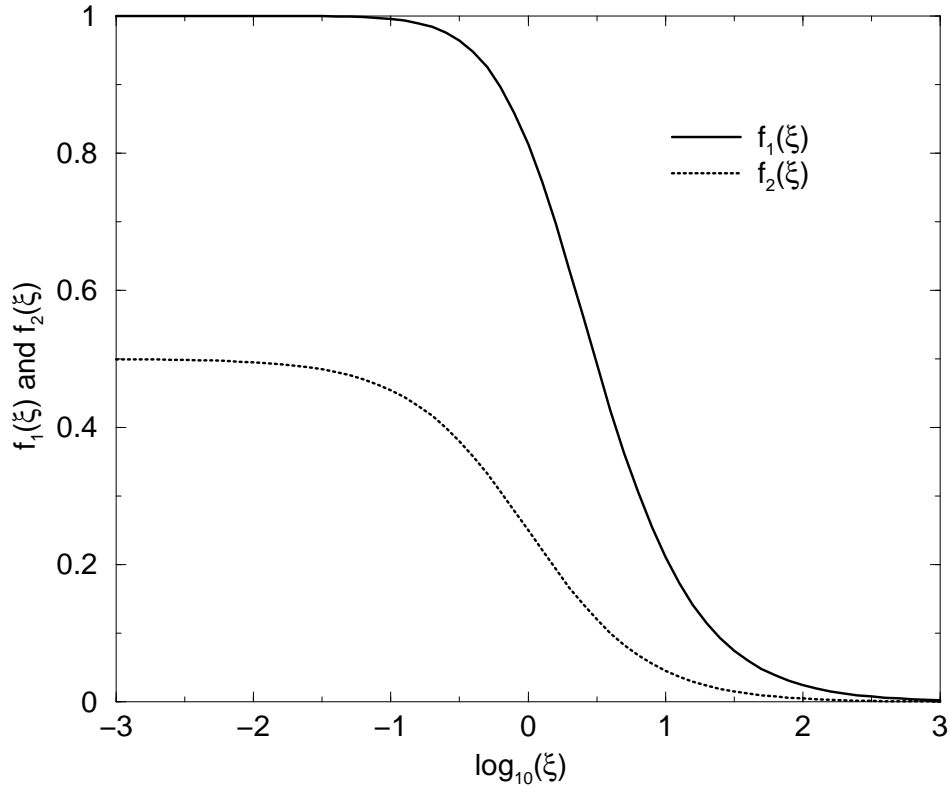,width=.7\textwidth}}
\caption{The functions $f_1\left(\xi\right)$ [Eq.~(\ref{2f1})]
(solid line) and $f_2\left(\xi\right)$ [Eq.~(\ref{2f2})] (dotted line)
plotted vs.~$\log_{10}(\xi)$.}
\label{flucfig2}
\end{figure}
Carrying out the second integrations in Eq.~(\ref{f1}) and (\ref{f2})
we obtain
\be\label{2f1}
f_1\left(\xi\right)=\f{1}{\left(1+\xi\right)}+\f{\xi}{\left(1+\xi\right)^2}
+\f{\xi^2}{2\left(1+\xi\right)^3}
\ee
and
\be\label{2f2}
f_2\left(\xi\right)=\f{1}{2\left(1+\xi\right)}.
\ee
The integrations in the calculation of $f_1$ and $f_2$ above were again 
carried out using the calculus of residues and were checked by numerical
integration.
The functions $f_1\left(\xi\right)$ and $f_2\left(\xi\right)$ are
plotted in Fig.~\ref{flucfig2}.

We have thus shown that a complete description of the decay-out of a 
superdeformed band within the energy average approach requires three 
requires three dimensionless variables, 
$\Ga^\da/\Ga_S$, $\Ga_N/d$ and $\Ga_N/(\Ga_S+\Ga^\da)$. 
We find
\\ (1) The average contribution of the background to the intraband decay 
intensity, $I_\ina^\av$, [Eq.~(\ref{2Iav})] depends only on $\Ga^\da/\Ga_S$. 
\\ (2) The average of the fluctuation contribution,
$\ov{I_\ina^\fl}$, [Eq.~(\ref{2ovIfl})]
depends on two variables: $\Ga^\da/\Ga_S$ and $\Ga_N/d$.
\\ (3) The variance, $\ov{\left(\Delta I_\ina\right)^2}$, 
[Eq.~(\ref{3var})] depends on three variables:
$\Ga^\da/\Ga_S$, $\Ga_N/d$ and $\Ga_N/(\Ga_S+\Ga^\da)$.
\begin{figure}[ht]
\centerline{\epsfig{figure=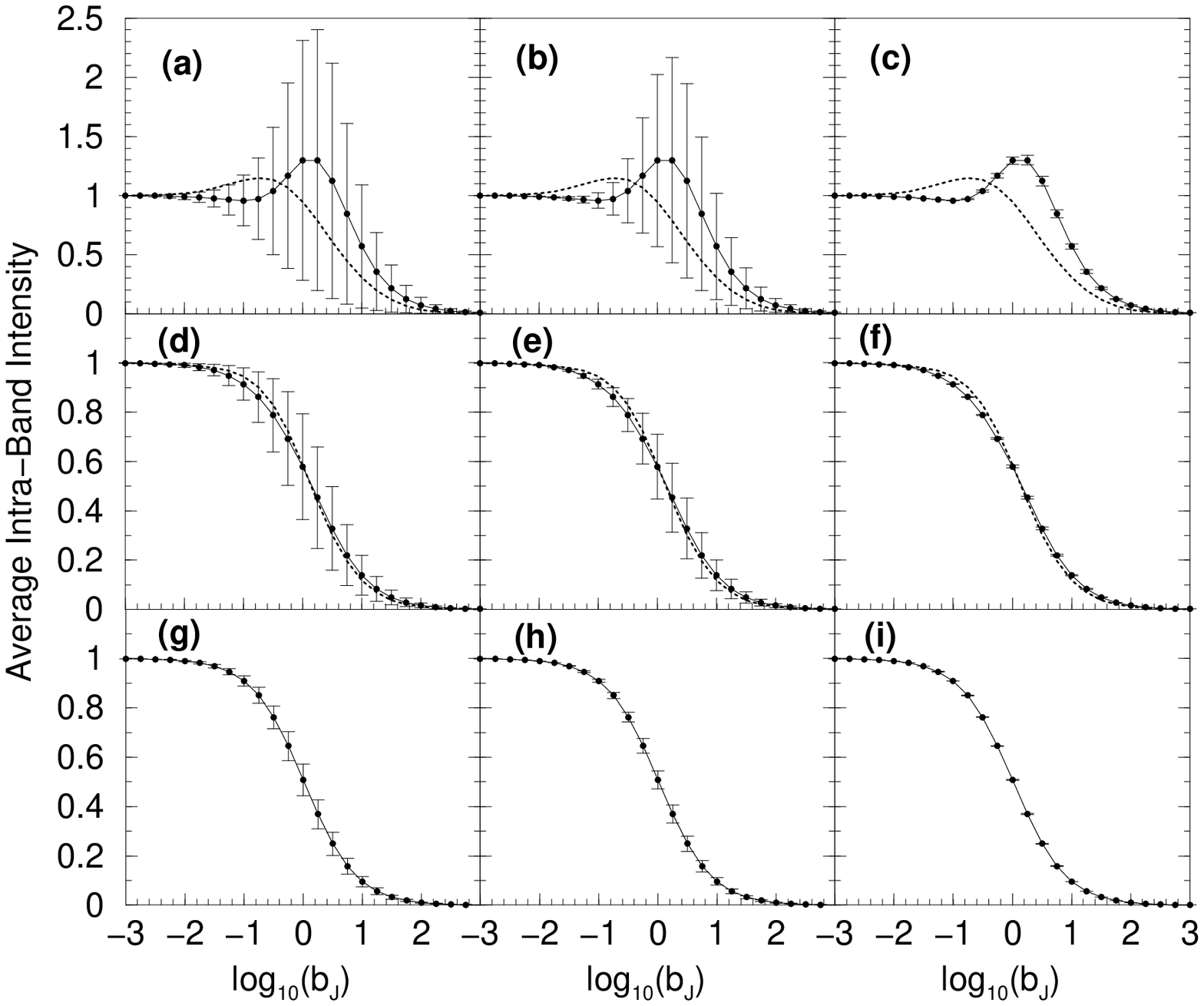,width=.7\textwidth}}
\caption{Average intraband intensity $\ov{I_\ina}$~vs.~$\log_{10}(b_J)$ 
where $b_J\equiv\Ga^\da/\Ga_S$. The filled circles
were calculated using Eq.~(\ref{ovI}) together with 
Eqs.~(\ref{2Iav}) and (\ref{2ovIfl}). The error bars for the filled
circles show $\sqrt{\ov{\left(\Delta I_\ina\right)^2}}$ calculated using 
Eq.~(\ref{3var}) with with $\xi$ in the form given by Eq.~(\ref{3xi}).
The dotted lines were calculate in the same manner as the filled circles
except that GW's fit formula, Eq.~(\ref{gufit}), was used in the place of 
Eq.~(\ref{2ovIfl}). 
The variable $\Ga_N/d$ took the following values: 
0.1 in graphs (a), (b) and (c);
1 in graphs (d), (e) and (f); 
10 in graphs (g), (h) and (i).
The variable $\Ga^\da/\Ga_N$ took the following values: 
$10^{-3}$ in graphs (a), (d) and (g);
1 in graphs (b), (e) and (h); 
$10^{3}$ in graphs (c), (f) and (i).}
\label{flucfig3}
\end{figure}

Fig.~\ref{flucfig3} shows a plot of the average intraband decay 
intensity $\ov{I_\ina}$ [Eq.~(\ref{ovI})] calculated using Eqs.~(\ref{2Iav})
and (\ref{2ovIfl}) for $I_\ina^\av$ and $\ov{I_\ina^\fl}$.
For comparison we also show the $\ov{I_\ina}$ which results when the GW 
fit formula [Eq.~(\ref{gufit})] is used for $\ov{I_\ina^\fl}$ instead of 
Eq.~(\ref{2ovIfl}). The two curves are barely distinguishable for $\Ga_N/d=1$.
The GW fit formula incorrectly gives intensities which 
are greater than unity when $\Ga_N/d=0.1$ as does our Eq.~(\ref{2ovIfl}).
The exact formula of GW [Eq.~(24) in GW] does not suffer from this problem.
Our results are only strictly valid when $\Ga_N/d\gg 1$.

The error bars in Fig.~\ref{flucfig3} were calculated using Eq.~(\ref{3var})
with $\xi$ in the form given by Eq.~(\ref{3xi}). We calculate 
$\sqrt{\ov{\left(\Delta I_\ina\right)^2}}$ for three values of the ratio 
$\Ga^\da/\Ga_N$ (see figure caption). Since 
$\ov{\left(\Delta I_\ina\right)^2}$ is proportional to $I_\ina^\fl$
it also increases monotonically to a maximum before decrease monotonically
to zero as a function of $\Ga^\da/\Ga_S$. For the same reason the error 
increases monotonically with decreasing $\Ga_N/d$. The error estimate 
presented in GW exhibits the same trends with $\Ga^\da/\Ga_S$ and $\Ga_N/d$.

\begin{figure}[ht]
\centerline{\epsfig{figure=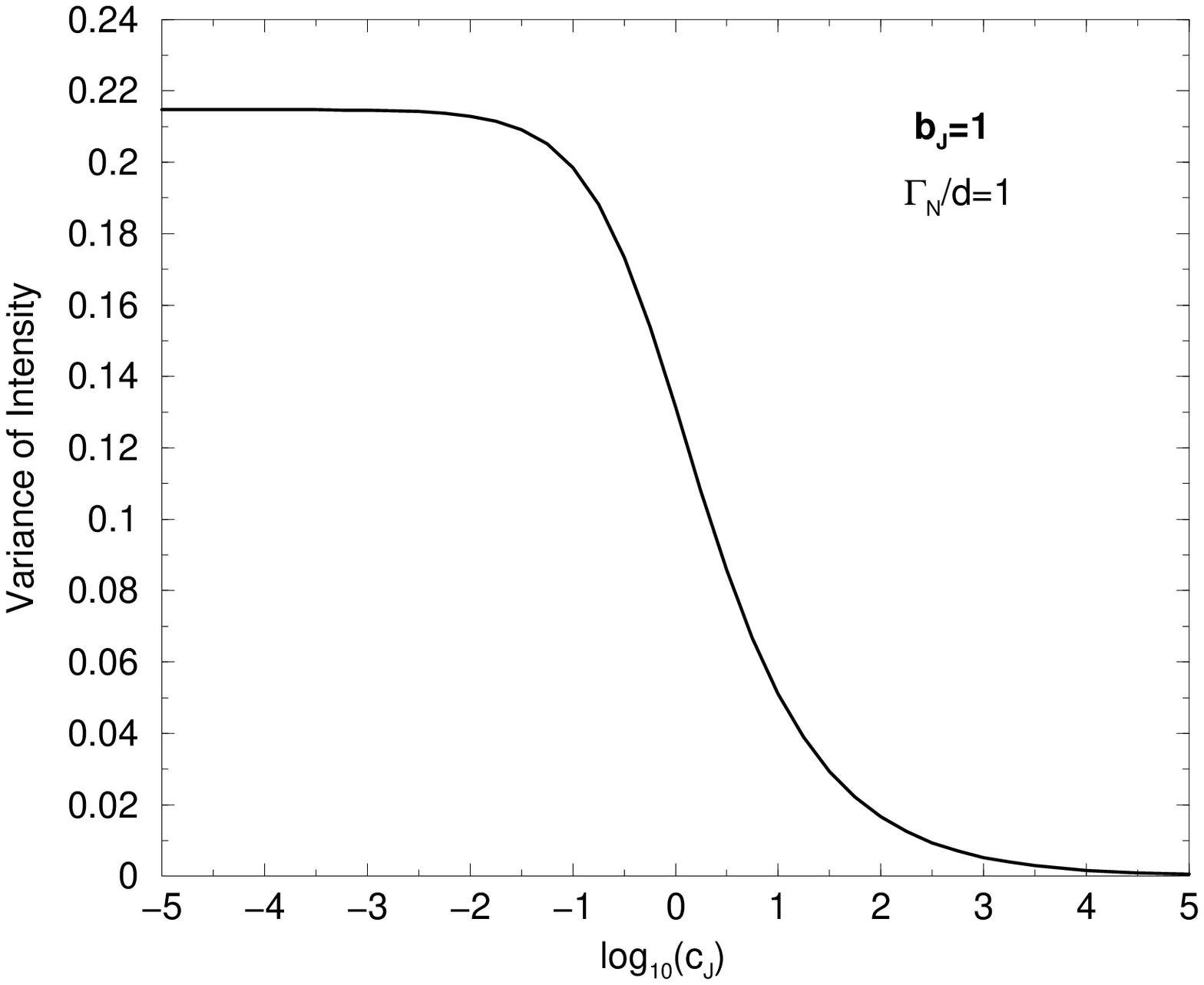,width=.7\textwidth}}
\caption{The standard deviation of the decay intensity 
$\sqrt{\ov{\left(\Delta I_\ina\right)^2}}$~vs.~$\log_{10}(c_J)$ 
where $c_J\equiv \Ga^\da/\Ga_N$ plotted using Eq.~(\ref{3var}) with $\xi$ in 
the form given by Eq.~(\ref{3xi}) for fixed $b_J=\Ga^\da/\Ga_S$ and $\Ga_N/d$.}
\label{flucfig4}
\end{figure}

Since the variance depends only on $(\Ga_S+\Ga^\da)/\Ga_N$ in addition to
$\Ga^\da/\Ga_S$ and $\Ga_N/d$, upon fixing the latter two variables
the variance may be considered a function of any {\em one} of $\Ga^\da/\Ga_N$,
$\Ga_S/\Ga_N$, $\Ga^\da/d$ or $\Ga_S/d$ [see Eqs.~(\ref{2xi}) and
(\ref{3xi})]. In Fig.~\ref{flucfig3} we chose to use 
Eq.~(\ref{3xi}) for $\xi$ and fixed the value of $\Ga^\da/\Ga_N$. When 
$\Ga_S/\Ga_N$ is fixed instead, a slightly different dependence on
$\Ga^\da/\Ga_S$ is obtained [compare Eqs.~(\ref{2xi}) and (\ref{3xi})].
Fig.~\ref{flucfig4} shows a plot of the standard deviation,
$\sqrt{\ov{\left(\Delta I_\ina\right)^2}}$, [Eq.~(\ref{3var})] as a 
function of $\Ga^\da/\Ga_N$ for fixed $\Ga^\da/\Ga_S$ and $\Ga_N/d$.
Ultimately, the variance like the intensity is a function of the spin of the
decaying nucleus and could provide an additional probe to the spin 
dependence of the barrier separating the SD and ND wells which is contained
in the spreading width $\Ga^\da$ \cite{Vi 90,Sh 92,Sh 93,Sh 00,Yo 00}. 

Our result for the variance of the decay intensity, 
$\ov{\left(\Delta I_\ina\right)^2}$ [Eq.~(\ref{3var})] has a structure 
reminiscent of Ericson's expression for the variance of the cross section 
\cite{Er 63}. This connection will be fully explored in the 
Section~\ref{comp}. For now we note only that what distinguishes 
Eq.~(\ref{3var}) from Ericson's expression for the variance of the cross 
section are the functions $f_1\left(\xi\right)$ and $f_2\left(\xi\right)$ 
which result from the energy integrations in Eq.~(\ref{2var}). 
\section{Variance of the decay intensity versus autocorrelation 
functions of statistical nuclear reaction theory}
\label{comp}
\subsection{Results familiar from statistical nuclear reaction theory}
\label{reaction}
All moments of the $S$-matrix,  $S_{ab}(E)$, the 
quantities that describe the way in which $S_{ab}(E)$ fluctuates about it's 
average, $\ov{S_{ab}(E)}$, can be expressed in terms of $\ov{S_{ab}(E)}$ 
itself \cite{Ri 75}. Normally, specific 
moments such as the amplitude and cross section autocorrelation functions
are expressed in terms of transmission coefficients, defined to be 
\be
T_a=T_{aa}, \hspace{.5cm} 0\le T_a \le 1,
\label{Ta}
\ee
and their generalisation
\be
T_{ab}=1-\sum_c\ov{S_{ac}}\hspace{1mm}\ov{{S_{bc}}^*}
=\sum_c\ov{S^\fl_{ac}{S^\fl_{bc}}^*}.
\label{Tab}
\ee
Here $S^\fl_{ab}=S_{ab}-\ov{S_{ab}}$ is the fluctuating part of the
$S$-matrix. The transmission coefficient $T_a$ is the probability 
of transmission from a compound nucleus resonance to channel $a$ and is 
obtained from the optical model.

In what follows we quote results for both the amplitude and cross section 
autocorrelation functions in the overlapping resonance region for the 
purpose of comparison with our results for the decay intensity.
The amplitude autocorrelation function $c_{ab}(E,E')$ for amplitude 
\be
\label{3A}
A_{ab}=\delta_{ab}-iS_{ab},
\ee  
is defined by
\bea\label{c}
c_{ab}(E,E')&=&\ov{A_{ab}(E){A_{ab}(E')}^*}
-\ov{A_{ab}(E)}\hspace{1mm}\ov{{A_{ab}(E')}^*}
\\\label{2c}&=&\ov{A^\fl_{ab}(E){A^\fl_{ab}(E')}^*}
\\\label{3c}&\approx&\ov{\sig^\fl_{ab}}\f{i\Ga}{E-E'+i\Ga}.
\eea
The correlation width $\Ga$ is given by
\be
\Ga=\f{d}{2\pi}\sum_a T_a.
\label{GaT}
\ee
Derivation of Eq.~(\ref{3c}) requires the assumption that 
$\ov{A_{ab}(E)}=\ov{A_{ab}(E')}$. This assumption together with 
statistical assumptions equivalent to those employed in our 
treatment of the factor $a(\ve)$ [Eq.~(\ref{a})] gives rise to a 
$c_{ab}(E,E')$ which depends solely on the difference of the two energies 
$E-E'$. 

Eq.~(\ref{2Aflauto}), which is essentially $c_{00}(E,E')$ does not
depend solely on $E-E'$. The background energy modulation in both $E$ and $E'$
which is characteristic of an isolated doorway resonance is explicit. 
In the present case it cannot be assumed that 
$\ov{A_{00}(E)}\ne\ov{A_{00}(E')}$ for arbitrary $E$ and $E'$.
The double integral for the variance of the decay intensity
[Eq~(\ref{2var}) and Eq.~(\ref{varC}) below] is sensitive to this fact this
as it contains products of the these background amplitudes at arbitrary 
$E$ and $E'$.

The amplitude autocorrelation function, Eq.~(\ref{2Aflauto}), contains two
distinct energy dependences, one characterised by $\Ga_N$ which is 
analogous to Ericson's correlation width as defined by Eq.~(\ref{GaT})
and another characterised by $\Ga_S+\Ga^\da$ the width of the doorway. 
Writing Eq.~(\ref{2Aflauto}) in terms of $x$ and $x'$ defined by 
Eqs.~(\ref{x}) and (\ref{x'})
\be
c_{00}(E,E')=\f{16i}{\xi}I_\ina^\av\ov{I_\ina^\fl}
\f{1}{\left[x+i\right]^2
\left[x-x'+2i/\xi\right]
\left[x'-i\right]^2},
\ee
we see that it in fact depends only on the ratio $\Ga_N/(\Ga_S+\Ga^\da)$ and
it is through Eq.~(\ref{2Aflauto}) that this variable enters our 
calculation of the variance of the decay intensity. 

The amplitude autocorrelation function is not an observable quantity. The
correlation width, Eq.~(\ref{GaT}), must be extracted from correlation 
analysis of the cross section. The cross section autocorrelation function, 
$C_{ab}(E,E')$, for cross section $\sig_{ab}=|A_{ab}|^2$ is defined by 
\bea
C_{ab}(E,E')&=&\ov{\sig_{ab}(E)\sig_{ab}(E')}
-\ov{\sig_{ab}(E)}\hspace{2mm}\ov{\sig_{ab}(E')}.
\label{C}
\\\label{1C}&\approx&|c_{ab}(E,E')|^2+2\mbox{Re}\hspace{1mm}
\ov{{A_{ab}(E)}^*}c_{ab}(E,E')\ov{A_{ab}(E')}
\\&\approx&\left[{\ov{\sig^\fl_{ab}}}^2
+2\sig^\av_{ab}\ov{\sig^\fl_{ab}}\right]
\f{\Ga^2}{(E-E')^2+\Ga^2}.
\label{2C} 
\eea
Here, $\sig^\av_{ab}=\left|\ov{A_{ab}}\right|^2$ is the background cross
section. The fluctuation contribution to the cross section in terms of the 
transmission coefficients is given by the Hauser-Feshbach formula
\be
\sig^\fl_{ab}\approx\f{T_aT_b}{\sum_c T_c},
\ee 
or some modification of it designed to account of
width fluctuations, direct reactions etc. \cite{Fe 92,Ve 86}. 

Eq.~(\ref{2var}) for the variance of the decay intensity can be written 
in terms of thecross section autocorrelation function defined by 
Eq.~(\ref{1C}) as
\be
\ov{\left(\Delta I_\ina\right)^2}
=\left(2\pi\Ga_S\right)^{-2}
\int_{-\infty}^{\infty}dE\int_{-\infty}^{\infty}dE'C_{00}(E,E').
\label{varC}
\ee
The same comments concerning the energy independence of the background
amplitude apply to the derivation of Eq.~(\ref{2C}) as apply to the 
derivation of Eq.~(\ref{3c}). Likewise $C_{00}(E,E')$ in the case of the 
present paper [the integrand in Eq.~(\ref{2var})] is distinguished from 
Eq.~(\ref{2C}) by
it's explicit inclusion of the energy dependence of the background. 
Eqs.~(\ref{2var}) and (\ref{1C}) assume that that only pairwise correlations 
are present. Both Eqs.~(\ref{3c}) and (\ref{2C}) are valid when 
$\sum_a T_a\gg 1$, that is, in the strongly overlapping resonance region. 
\subsection{Expression of the decay intensity and variance in terms
of transmission coefficients}
Following GW we introduce two transmission coefficients, $T_0(E)$ and $T_N$,
where
\bea
T_0(E)=1-\left|\ov{S_{00}}\right|^2
&=&\f{\Ga_S\Ga^\da}{(E-E_0)^2+(\Ga_S+\Ga^\da)/4}.
\label{T0}
\\&=&\f{4I_\ina^\av(1-I_\ina^\av)}{4(E-E_0)^2/(\Ga_S+\Ga^\da)+1},
\label{1T0}
\eea
describes transmission from the $|Q\ket$ to the SD band and  
\be
T_N=2\pi \Ga_N/d,
\label{TN}
\ee
describes their transmission to ND states of lower spin. We have not derived
Eq.~(\ref{TN}). For the purposes of the present paper it can be taken as 
the definition of $T_N$. The reader is referred to the discussion in 
Section~VIII.H of  Ref.~\cite{Br 81} which contrasts the relation of
the correlation width, $\Ga$, to transmission coefficients with the 
the corresponding relation for the average width, $\ov{\Ga_q}$.

We have written $T_0(E)$
in the form given by Eq.~(\ref{1T0}) in order to emphasise that it is {\em not}
simply a function of a single dimensionless variable, the ratio
$\Ga^\da/\Ga_S$. It is energy dependent, the energy dependence being
characterised by $\Ga_S+\Ga^\da$, the total width of doorway state $|0\ket$.
Only it's maximum $T_0(E_0)=4I_\ina^\av(1-I_\ina^\av)$ 
can be expressed solely in terms of $\Ga^\da/\Ga_S$. Thus, a quantity
sensitive to the gross energy dependence of $T_0(E)$ should depend on 
$\Ga_S+\Ga^\da$. Writing the average decay intensity 
Eq.~(\ref{ovI}) in terms transmission coefficients
\be
\ov{I_\ina}=1-(2\pi \Ga_S)^{-1}\int_{-\infty}^{\infty}dE
\left\{T_0(E)-2\left[T_0(E)\right]^2/T_N\right\}, 
\label{3ovI}
\ee
we see that it
compares the total width of the doorway $|0\ket$ with the width 
for the feeding of $|0\ket$ [thanks to inclusion of the normalisation 
factor $2\pi\Ga_S$ in the definition of $I_\ina$ in Eq.~(\ref{I})].
The variance, Eq.~(\ref{2var}), may be written
\be\ov{\left(\Delta I_\ina\right)^2}
\approx\left(2\pi\Ga_S\right)^{-2}
\int_{-\infty}^{\infty}dE\int_{-\infty}^{\infty}dE'\left\{
\f{4{T_0(E)}^2{T_0(E')}^2}{\left[2\pi(E-E')/d\right]^2+T_N^2}
+4\mbox{Im}\f{\ov{{A_{00}(E)}}T_0(E)T_0(E')\ov{{A_{00}(E')}^*}}
{2\pi(E-E')/d+iT_N}\right\}.
\label{4var}
\ee
As discussed in Subsection~\ref{reaction} and made explicit by 
Eqs.~(\ref{1f1}) and (\ref{1f2}), the integrand 
of Eq.~(\ref{4var}) which clearly contains two characteristic energy scales in 
fact only depends on their ratio, $\Ga_N/(\Ga_S+\Ga^\da)$, the ratio of the 
correlation width to the doorway width. 

Eq.~(\ref{2Ga}) may also be expressed in terms of the transmission 
coefficients $T_0(E)$ and $T_N$. Using Eq.~(\ref{Gaup}) for $\Ga^\up$ we get
\be
\Ga=\f{d}{2\pi}T_N\left[1-\f{T_0(E)/T_N}{I_\ina^\av}\right],
\ee
so that the neglect of $\Ga^\up$ in $\Ga$ is justified when 
$T_0(E)/T_N\ll I_\ina^\av\le 1$. Let us also write 
the correlation length $\xi$, Eq.~(\ref{xi}), in terms of the transmission 
coefficients:
\be
\xi=(2\pi/T_N)\f{\Ga_S+\Ga^\da}{d}.
\ee

In the case compound nucleus scattering, extraction of $\Ga$ from a 
measurement of cross section autocorrelation function, using say
Eq.~(\ref{2C}), permits the determination of the density of compound nucleus 
states, $1/d$, by application of Eq.~(\ref{GaT}) \cite{Er 66}. 
A more recent example of energy-autocorrelation analysis
of may be found in Ref.~\cite{Papa:zk} where fluctuations in dissipative
binary heavy ion collisions are studied.
In the present case of the decay out of a superdeformed band 
extraction of $\xi$ from thevariance of the intensity, permits the 
determination of the ratio $(\Ga_S+\Ga^\da)/\Ga_N$, or, given $T_N$ 
(equivalently $\Ga_N/d$) determination of the ratio $(\Ga_S+\Ga^\da)/d$.
\subsection{Comparison with the results of Gu and Weidenm\"uller}
\label{gu}
GW also take inspiration from statistical nuclear reaction 
theory but use the MPI approach \cite{Verbaarschot:jn}. 
The MPI approach is concerned with the analytic calculation of  
ensemble averages, a procedure which is equivalent to the calculation 
of energy averages. Ref.~\cite{Verbaarschot:jn} use the supersymmetry method
of calculating ensemble averages to derive 
an exact expression for $\ov{S^\fl_{ab}(E) S^\fl_{cd}(E')}$. Their result is 
found to be expressible in terms of the difference of the two energies, $E-E'$,
and transmission coefficients. The transmission coefficients themselves are
expressed as functions of $(E+E')/2$.
The relationship between the results of \cite{Fe 92,Ka 73,Er 63} 
and those of \cite{Verbaarschot:jn} is discussed in Refs.~\cite{Yo 90,Ve 86}.
Several results of \cite{Fe 92,Ka 73,Er 63} can be obtained from that 
of Ref.~\cite{Verbaarschot:jn} by expanding in powers of the transmission 
coefficients or inverse powers of the sum of the transmission coefficients 
\cite{Ve 86}.

Calculation of the average of the fluctuation intensity requires the
energy integral of the average of the product of two $S$-matrix elements at 
the same energy. GW use the results of Ref.~\cite{Verbaarschot:jn} for 
$\ov{|S^\fl_{00}(E)|^2}$ to calculate the average decay intensity.
As was already noted in Section~\ref{intensity},
GW include the energy dependence of the background amplitude characteristic of
an isolated doorway resonance in their calculation by using the energy 
dependent transmission coefficient $T_0(E)$, Eq.~(\ref{T0}), in their 
Eq.~(24) for $\ov{I_\ina^\fl}$. The fact that we use the same energy 
dependence as GW for the background is responsible for the agreement we 
obtain with GW concerning $\ov{I_\ina}$'s dependence on $\Ga^\da/\Ga_S$.
The differences between our results and those of GW for the decay 
intensity stem from the assumptions we make which restrict our 
results to $\Ga_N/d\gg 1$.  

Calculation of the variance of the intensity requires the 
4-point function at two energies integrated over both energies, that is, 
it requires
$\ov{S_{00}^\fl(E){S_{00}^\fl(E)}^*S_{00}^\fl(E'){S_{00}^\fl(E')}^*}$
integrated over $E$ and $E'$. Calculation of the 4-point function at
two energies was carried out using the supersymmetry method in 
Ref.~\cite{Bo 89}. Their result, 
like that of Ref.~\cite{Verbaarschot:jn} for the 2-point function
depends explicitly only on $E-E'$ and the transmission coefficients
which are again expressed as functions of $(E+E')/2$. 
Within the assumption that only pairwise correlations are important,
as was assumed in Eqs.~(\ref{2var}) and (\ref{1C}), the 2-point function 
is enough to calculate the variance.
Ref.~\cite{Ve 86} showed numerically that the exact expression of 
Ref.~\cite{Verbaarschot:jn} specialised to the amplitude autocorrelation
function confirms the correctness of Eq.~(\ref{3c}) 
in the region of strongly overlapping resonances. 
However, unlike Eq.~(\ref{3c}), the amplitude autocorrelation function as 
given by Eq.~(\ref{2Aflauto}) depends on the background amplitude at two 
different energies,
that is, it depends on $\ov{A_{00}(E)}$ and $\ov{A_{00}(E')}$.
When $E=E'$ it reduces to Eq.~(\ref{Aflsqu}) which can be expressed in 
terms of the transmission coefficients $T_0(E)$ and $T_N$. Thus the decay 
intensity can be expressed in terms of these transmission coefficients as 
was done in Eq.~(\ref{3ovI}). The applicability of  
Ref.~\cite{Verbaarschot:jn} to calculation of the decay intensity 
owes itself to the fact that the decay intensity may be 
expressed in terms of transmission coefficients.

Eq.~(\ref{2Aflauto}) cannot be written in terms of $T_0([E+E']/2)$ and the
same applies to the variance as is apparent from Eq.~(\ref{4var}).
Thus it is not clear whether Ref.~\cite{Verbaarschot:jn} 
serves as a means to obtain results corresponding to 
Eqs.~(\ref{2Aflauto}) and (\ref{4var}) which are valid for arbitrary $\Ga_N/d$.
It would be an interesting challenge to derive an expression for the variance 
which could be used for any value of $\Ga_N/d$ since for the regions which 
have been most frequently studied experimentally \cite{Lo 00}, 
the $A\approx 150$ and $A\approx 190$ regions, $\Ga_N/d\ll 1$.  
 
GW do not use the supersymmetry method to calculate the variance. 
They instead estimate the variance by performing a numerical simulation. 
The analytic structure of the variance was not investigated in GW and their 
results make no reference to the variable $\Ga_N/(\Ga_S+\Ga^\da)$. 
Given the close resemblance of the
conclusions about the analytic structure of the decay intensity which may be 
inferred from the exact result of GW and our approximate result for 
$\Ga_N/d\gg 1$ it seems probable that the dependence of the variance on 
$\Ga_N/(\Ga_S+\Ga^\da)$ which we have found for $\Ga_N/d\gg 1$ persists for 
arbitrary $\Ga_N/d$.
\section{Conclusions}
\label{conclusions}
In conclusion, we have derived analytic formulae for the energy average 
and variance of the intraband decay intensity of a superdeformed band
in terms of variables which usefully describe the decay-out. 
The formulae given by Eq.~(\ref{2ovIfl}) for the fluctuation  
contribution to the average intensity $\ov{I_\ina^\fl}$ and by 
Eq.~(\ref{3var}) for the variance $\ov{\left(\Delta I_\ina\right)^2}$ were 
derived by making assumptions and approximations which are strictly valid 
only in the strongly overlapping resonance region, $\Ga_N/d\gg 1$. However, 
these formulae are seen from Figs.~\ref{flucfig2} and \ref{flucfig3} to work 
well when $\Ga_N/d$=1 and provide a qualitative description even when 
$\Ga_N/d$=0.1. This means that Eq.~(\ref{2ovIfl}) 
and Eq.~(\ref{3var}) cannot be applied to the mass 150 and 190 regions 
where $\Ga_N/d\sim 0.001$ but they may prove themselves of practical use 
in other 
mass regions. In any case our results clarify the analytic structure of the 
results obtained by GW. In particular we have 
revealed that the variance of the decay intensity depends on the correlation 
length $\Ga_N/(\Ga_S+\Ga^\da)$ in addition to the two dimensionless variables
$\Ga^\da/\Ga_S$ and $\Ga_N/d$ on which the average of the decay intensity
depends. Measurement of the variance of the decay intensity could yield the 
mean level density of the ND states in analogy with autocorrelation analysis 
of cross sections.

This work was partially supported by FAPESP.

\end{document}